%% file: main.tex
\documentclass{egpubl}
\usepackage{sca2022}

\usepackage{xcolor}
\usepackage{amssymb}
\usepackage{amsmath}
\usepackage{multicol}
\usepackage[utf8]{inputenc}
\usepackage{lipsum}  
\usepackage[normalem]{ulem}

\usepackage{algorithm,algpseudocode}
\algnewcommand{\To}{\textbf{To }}
\algnewcommand\Input{\item[\textbf{Input:}]}%
\algnewcommand\Output{\item[\textbf{Output:}]}%
\algnewcommand\algorithmicforeach{\textbf{for each}}
\algdef{S}[FOR]{ForEach}[1]{\algorithmicforeach\ #1\ \algorithmicdo}

\usepackage[T1]{fontenc}
\usepackage[english]{babel}
\usepackage[utf8]{inputenc}
\usepackage{booktabs, tabularx}

\usepackage{blindtext}

\usepackage{bm}
\usepackage{makecell}

\DeclareMathOperator*{\argmin}{arg\,min}

\newcommand{\ea}[1]{\textcolor{cyan}{#1}}

\newcommand{\drq}[1]{\textcolor{blue}{\textbf{[DR: #1]}}}
\newcommand{\mpc}[1]{{\textcolor{purple}{#1}}}
\newcommand{\MPC}[1]{{\textcolor{purple}{\textbf{[MP: #1]}}}}
\newcommand{\changes}[1]{\textcolor{red}{#1}}

\newcommand{\EAQ}[1]{{\textcolor{cyan}{\textbf{[EA: #1]}}}}

\SpecialIssuePaper         

\usepackage[T1]{fontenc}
\usepackage{dfadobe}  

\usepackage{cite}  
\BibtexOrBiblatex
\electronicVersion
\PrintedOrElectronic

\ifpdf \usepackage[pdftex]{graphicx} \pdfcompresslevel=9
\else \usepackage[dvips]{graphicx} \fi

\usepackage{egweblnk} 

\usepackage[switch]{lineno}

\title[Generating Upper-Body Motion for Real-Time Characters Making their way through Complex Environments]{Generating Upper-Body Motion for Real-Time Characters \\ Making their Way through Dynamic Environments}

\author[E. Alvarado et al.]
{\parbox{\textwidth}{\centering Eduardo Alvarado$^{1}$, Damien Rohmer$^{1}$, Marie-Paule Cani$^{1}$}
\\
{\parbox{\textwidth}{\centering 
 \{eduardo.alvarado-pinero,damien.rohmer,marie-paule.cani\}@polytechnique.edu\\
$^1$ LIX, École Polytechnique/CNRS, Institut Polytechnique de Paris, Palaiseau, France
}}}
      
\usepackage{tikz}

\tikzset{
    vertex/.style = {
        circle,
        fill            = black,
        outer sep = 2pt,
        inner sep = 1pt,
    }
}

\begin{document}


\teaser{
    \includegraphics[width=\linewidth]{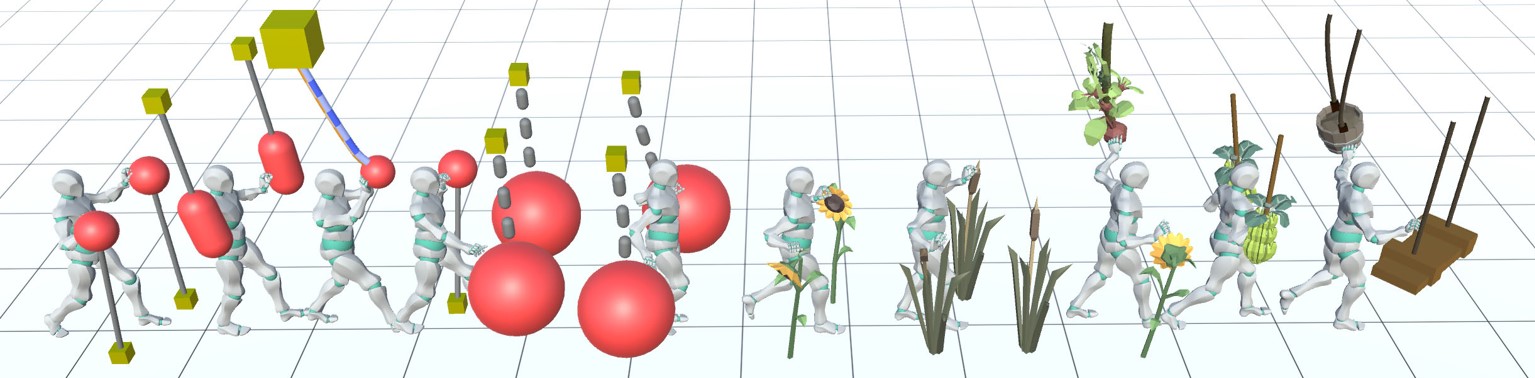}
    \centering
    \caption{
    Thanks to a hybrid physical-kinematic model allowing the adjustment of tension/relaxation in the upper-body, our responsive character is able to anticipate interactions in order to make its way through a variety of obstacles, while being driven by an input walking motion. 
    }
    \label{fig:teaser}
}

\maketitle
\begin{abstract}
Real-time character animation in 
dynamic environments requires the generation of plausible upper-body movements regardless of the nature of the environment, including non-rigid obstacles such as vegetation.
We propose a flexible model 
for upper-body interactions, based on the anticipation 
of the character's surroundings, and on antagonistic controllers 
to adapt the amount of muscular stiffness and response time to better deal with obstacles.
Our solution relies on a hybrid method for character animation that couples 
a keyframe sequence with kinematic constraints and lightweight physics. 
The dynamic response of the character's upper-limbs leverages antagonistic controllers, allowing us to tune tension/relaxation in the upper-body without diverging from the reference keyframe motion.
A new sight model, 
controlled by procedural rules, enables high-level authoring of the way 
the character generates interactions by adapting its stiffness and reaction time.
As results show, our real-time method offers precise and explicit control over the character's behavior and style, while seamlessly adapting to new situations. 
Our model is therefore well suited for gaming applications.


\end{abstract}  

\input{Text/01-Introduction}
\input{Text/02-Related-Work}
\input{Text/04-Hybrid-Model}
\input{Text/05-Antagonistic-Control}
\input{Text/06-Anticipation-System}

\input{Text/07-Results}
\input{Text/09-Conclusion}
\input{Text/99-Acknowledgments}

\bibliographystyle{eg-alpha} 
\bibliography{Text/bibliography}       



\input{Text/10-Appendix}

\end{document}

%% file: Text/01-Introduction.tex
\section{Introduction}
\label{introduction}

Being able to animate virtual characters navigating through complex, dynamic environments is of utmost importance for video games and virtual reality applications.
Natural scenes that typically include 
tall plants, bushes, and trees of various types and sizes are particularly challenging.

In real life, 
we, humans, constantly anticipate and adapt our
upper-body 
motions when making our way through
complex environments. Roughly evaluating the expected stiffness and dynamics 
of 
obstacles helps us anticipate interactions, and in particular tune our muscular stiffness 
and reaction speed in order to efficiently push them out of the way or avoid their trajectories. 
For example, we stay relaxed making space around us in the middle of tall grass, but we tense our muscles to fend off a stiffer branch of a tree or quickly protect our head when a potentially dangerous object comes toward us.
Moreover, the way we interact with obstacles does not only depend on our prediction but also adapts dynamically to the environment's response, especially if it differs from 
our expectations.

While recent video games produce highly realistic images of natural scenes~\cite{KojimaProductions2019, GuerrillaGames2022}, the range of dynamic interactions they can handle between a character and its environment remains quite limited. Indeed, the combined need of predictable overall motion and high performance usually prevents the use of fully physically-based models in gaming applications. Therefore, modeling interactions usually requires the dynamic selection of a best animation clip from a motion library, and coupling it in real time with a suitable response of the environment, as was done, for instance, to model characters walking on loose grounds such as mud or snow~\cite{RockstarStudios2018}.
%
%
Leveraging such lightweight animation clips with simple adaptation to various environments would tremendously ease the design of video games. Ideally, a game designer should be able to use standard pre-recorded animations, and refine and adapt these efficiently in a semi-automatic way to different behaviors. To this end, the gestures of the character should (i) automatically adapt in a plausible way to its surrounding environment when needed (i.e., a character making their way through dynamic environments such as in dense vegetation), (ii) still follow, while doing so, the high-level intent set by the designer, while (iii) relying on the initial animation clip for the general motion.

In this work, we propose a real-time animation model for the upper-body of human-like virtual characters; it provides high-level behavioral control in order to interact plausibly with dynamic obstacles. Composed of an anticipation mechanism based on sight, followed by 
a rule-based action module, our model makes the character responsive to its surroundings by allowing it to push away dynamic obstacles
made of hierarchical articulated rigid bodies associated to visual skinned-rigs%
, and make its way through the observed environment.
To achieve this, we introduce a hybrid, layered character model that couples the input
keyframe animation with a lightweight physics-inspired model, 
used to dynamically adapt the upper-body animation. The character is able to anticipate interactions based on high-level rules 
that take into account ray-cast visibility as well as the predicted stiffness of obstacles. This anticipation mechanism is used to guide an 
Inverse Kinematics (IK) objective 
that drives the character's gestures with suitable synchronization and amount of muscular stiffness. 
Our solution allows characters to react to any upcoming obstacle, and therefore to plan actions, through simple real-time 
query about their surroundings. The interaction process relies on a dedicated reformulation of 
Neff and Fiume's antagonist control method~\cite{Neff2002}, which enables us to control tension/relaxation 
during a character's motion without affecting
the position and angular objectives of its limbs.
Ultimately, our model generates plausible action for a character, whatever its current state 
and the nature of obstacles, while remaining loosely driven 
by the input kinematic motion clip.

\noindent
Our technical contributions are threefold:

\begin{itemize}
\item A real-time, hybrid character model coupling 
keyframe animations with a lightweight physics-inspired model for each upper-body limb separately. It makes the upper-body of kinematic-controlled characters able to interact through plausible dynamic responses (Sec.~\ref{hybrid_character_model}).
\item An extension of antagonist controllers~\cite{Neff2002} allowing the intuitive tuning of gestures and interaction styles by dissociating the position and orientation of the limbs from their degree of muscular rigidity (Sec.~\ref{antagonist_control}).
\item An efficient, yet generic and customizable anticipation mechanism
that enables our model to couple high-level procedural rules with metadata from observed obstacles. By driving the kinematic controllers and anticipating the amount of tension 
or reaction time required, this mechanism generates character animations that adapt to the surroundings (Sec.~\ref{anticipation}).
\end{itemize}
Fig.~\ref{fig:teaser} illustrates the application of our method to 
a real-time character making 
its way through a variety of 
dynamic environments, including deformable obstacles.

%% file: Text/02-Related-Work.tex
\section{Related Work}
\label{related_work}



Generating the reactive motion for an agent in a dynamic environment is an active research topic. It was addressed in different fields, from Computer Graphics (CG) to robotics and bio-mechanics (e.g.,~\cite{Shahabpoor2017}). For the sake of conciseness, we focus here on CG research tackling the reciprocal influence between character motion and animation of its virtual surroundings.




\subsection{Controlling Physically-Based Characters}


Since the early years of CG animation, physically-based models 
have been explored to represent reactive characters in dynamic environments~\cite{Raibert1991, Hodgins1995}. \emph{Ragdoll} models represent each limb as a constrained articulated rigid-body with prescribed mass and inertia. The character's motion can be handled by applying a set of actuator forces and torques to the rigid bodies coupled with a numerical time integrator. However, computing coherent forces over time to achieve a given motion is a complex problem. A popular approach relies on the use of high-level \emph{controllers} for joint actuation, providing a trade-off between motion plausibility and the complexity of user-control~\cite{Geijtenbeek2012}.
Proportional-Derivative (PD) controllers are in particular considered as common ground for character-animation methods~\cite{Yin2007, Wang2010}. 
However, despite their simplicity, instability and 
stiffness control are still recurrent problems when reproducing highly-dynamic and accurate animations. 

Stable Proportional-Derivative (SPD) controllers~\cite{Tan2011, Yin2020} 
introduce the idea of incorporating the next simulation step into the computation of forces, thus improving numerical stability and performance. 
Simplified physical models~\cite{Kwon2010, Kwon2017}, stochastic optimal control~\cite{Liu2010}, or Model Predictive Control (MPC)~\cite{Tassa2012, Tassa2014, Eom2019} have also been successfully used to animate virtual agents with PD controllers at their joints. In terms of motor parameterization, Abe et al.~\cite{Abe2004} 
add momentum constraints to generate more plausible physical 
movements. Proportional controllers mix stiffness 
(which models the appearance of tension/relaxation in the character) and the joint orientation at 
the equilibrium state. This dependence may 
hamper the intuitiveness and precision that a user 
may want to address when parameterizing its character behavior. 
Neff and Fiume~\cite{Neff2002} introduce an antagonistic-based PD formulation, allowing them to decouple stiffness and 
equilibrium-state orientation. In our work, we leverage and extend 
their antagonistic formulation for arbitrary 3D limb 
motions, and demonstrate that it can be used to dynamically edit 
keyframe animations in an online fashion and to decouple stiffness and position 
controls. 
To the best of our knowledge, joint-based stiffness control has shown benefits in robotics~\cite{Buchli2011, Bhattacharjee2015, Martin-Martin2019}, but has not yet been fully explored for character animation.


The use of Reinforcement Learning (RL) has become prevalent in recent years for optimizing physically-based controller parameters~\cite{Kwiatkowski2022}. These approaches can preserve 
character balance to achieve realistic locomotion up to complex acrobatic gestures~\cite{Chentanez2018, Peng2018, Peng2020}.
However, it is still a difficult task to define an accurate, yet general reward system that performs well in selecting the best action under a multitude of scenarios.
Thanks to the democratization and availability of motion capture data, Deep Learning (DL) based motion synthesis has been very successful in 
reproducing 
lifelike character 
motions for prescribed-scenarios~\cite{Holden2017, Holden2020, Mourot2021}. 
Most particularly, DL has been combined with RL to handle real-time dynamic simulated behaviors while preserving the naturalness of the training data. The effectiveness of such combination was demonstrated for characters maintaining their 
balance in new environments with moving solid obstacles~\cite{Bergamin2019, Wang2020}. However, learning-based approaches still suffer from limitations that make them hard to use for efficient game-like 
setups in 
natural environments. First, precise authoring of learned 
behaviors is highly indirect and hard to predict; yet, this aspect is of utmost importance for game design. Second, natural environments are characterized by diverse deformable elements such as various types of 
vegetations or uneven terrains. Each natural element may be associated with different physical properties and thus, could trigger specific character behaviors. Pre-training all possible deformations and behaviors would be, at best, very challenging. This is also orthogonal to the process of current game development, where tuning of flexible behaviors and insertion of new environment assets have to be as lightweight as possible.

\subsection{Hybrid Character Models using Kinematics}

Kinematic-driven virtual characters are extremely efficient to compute and allow intuitive formulations for constraints and objectives. Hybrid models combine physical or data-driven representations with kinematics to offer a trade-off between automatic motion quality, automatic pose adaptation, and user-control.
The use of space-time constraints~\cite{Witkin1988} is a common way to describe 
kinematics objectives while preserving character dynamics, but it involves an optimization procedure 
that is not applicable at run time.
The use of Inverse Kinematics (IK) is an intuitive representation to specify end-effector objectives in a kinematic-chain~\cite{Aristidou2017}. It was used, for instance, in combination with physical 
constraints~\cite{Boulic1996} 
and hierarchical motion curve editing~\cite{Lee1999}
; it was also combined with short-term dynamical effects~\cite{Rahgoshay2012}. Extending motion synthesis to the morphology of an arbitrary character was proposed in integrating procedural techniques with a gait generator in dynamic environments for both quadruped and multi-legged characters~\cite{Karim2012,Karim2012a}.
A lightweight physically-based model reduced to a single inverse pendulum was combined with 
keyframe animations to generate a responsive, real-time character for augmented reality applications~\cite{Mitake2009}.

Similar to our approach, some 
work proposed local hybrid models addressing the motion of some parts of the upper-body, such as the arms.
Zordan 
and Hodgins~\cite{Zordan1999} added contact constraints to locally modify motion capture data, and 
extended it further to integrate dynamic responses~\cite{Zordan2002}. Arm motion was also studied using learning-based approaches in sport applications~\cite{Liu2018}, and used to infer lower-body motions in VR~\cite{Yang2021}.
In our work, we introduce a framework that helps the user define upper-body animations by locally combining 
keyframes and light physical control.

\subsection{Controlling Characters in Natural Environments}

Natural scenes are characterized by a rich set of dynamic and deformable elements , which might affect how the character actively behaves. Although visuomotor systems have been used to adapt the character based on external observations~\cite{Eom2019}, generating plausible, yet general motions that remain controllable
for virtual characters interacting with such environments
-- in real-time --
is challenging, due to the computational cost of a full-scale physical simulation of both, characters and deformable materials
that might constitute the ground or vegetation.
Layered models embedding simplified representations of the environment's response have been successfully used for real-time applications. 
For instance, a simplified fluid representation was used for swimming characters~\cite{Yang2004}, or for windy environments~\cite{Lentine2011}. A simplified friction model was used to represent 
human locomotion efficiently on 
semi-flooded grounds~\cite{Bermudez2018} or through dense vegetation represented using billboards~\cite{Paliard2021}.
An essential aspect 
in plausible natural environments is the bi-directional interaction between the character and 
the scene. The character should not only adapt to 
its surroundings, but the environment itself should also be deformed by the character's actions.  
In the resulting two-way interaction, the character's behavior should dynamically adapt to the ongoing deformation. Such coupled interactions were proposed for lower-body 
motions on soft natural grounds, such as mud or snow~\cite{Alvarado2022}. 
However, to the best of our knowledge, no work has yet tackled the 
real-time action-reaction of the character's upper-body in interaction with a natural environment.

%% file: Text/04-Hybrid-Model.tex
\section{Hybrid Character Model for Upper-Body Interactions}
\label{hybrid_character_model}

We introduce a hybrid model that couples
keyframe animation and lightweight physical simulation for human-like characters. Our novel system aims to be flexible enough to coexist with current gaming animation pipelines, such as in \textsl{Unity 3D}, 
by bringing together hand-crafted animations or motion-capture data, with on-the-fly actions and reactions to new situations.
The hybrid model is able to switch between physical and kinematic spaces independently for the different body parts. This allows the game designer to improve the interactive motion from the predefined kinematic controllers, based on the effects that the dynamic, deforming environment might have upon specific body parts, as well as on the actions that the character physically exerts on its surroundings.

We define the character as a human (bipedal) model represented by a skin mesh, rigged to an articulated animation skeleton. 
This model can be associated to a
set of predefined animations such as walking or running, 
either manually defined by 
keyframes or, alternatively, from motion-capture data.
In this work, we call \textit{input skeleton} the animated skeleton that follows these predefined movements. It will be used as a soft constraint to guide our hybrid model.
In parallel, we associate a 
physically-based ragdoll model to a subset of the character's limbs
%
%
using an anchor system (that will be described later). 
Each \emph{physical} limb is described as a rigid-body model defined by its mass, its center of mass, and its inertia tensor. Limbs are connected by joints, with angular limits for each degree of freedom.
A physical simulator computes 
the 
movements of this skeleton model, and takes into account 
the various forces and torques acting on it, 
such as the action of the weight on each limb, any other external force such as response forces due to interactions, as well as the actuator torques computed by our approach to animate the skeleton.

\begin{figure}[htb]
  \centering
  \includegraphics[width=\linewidth]{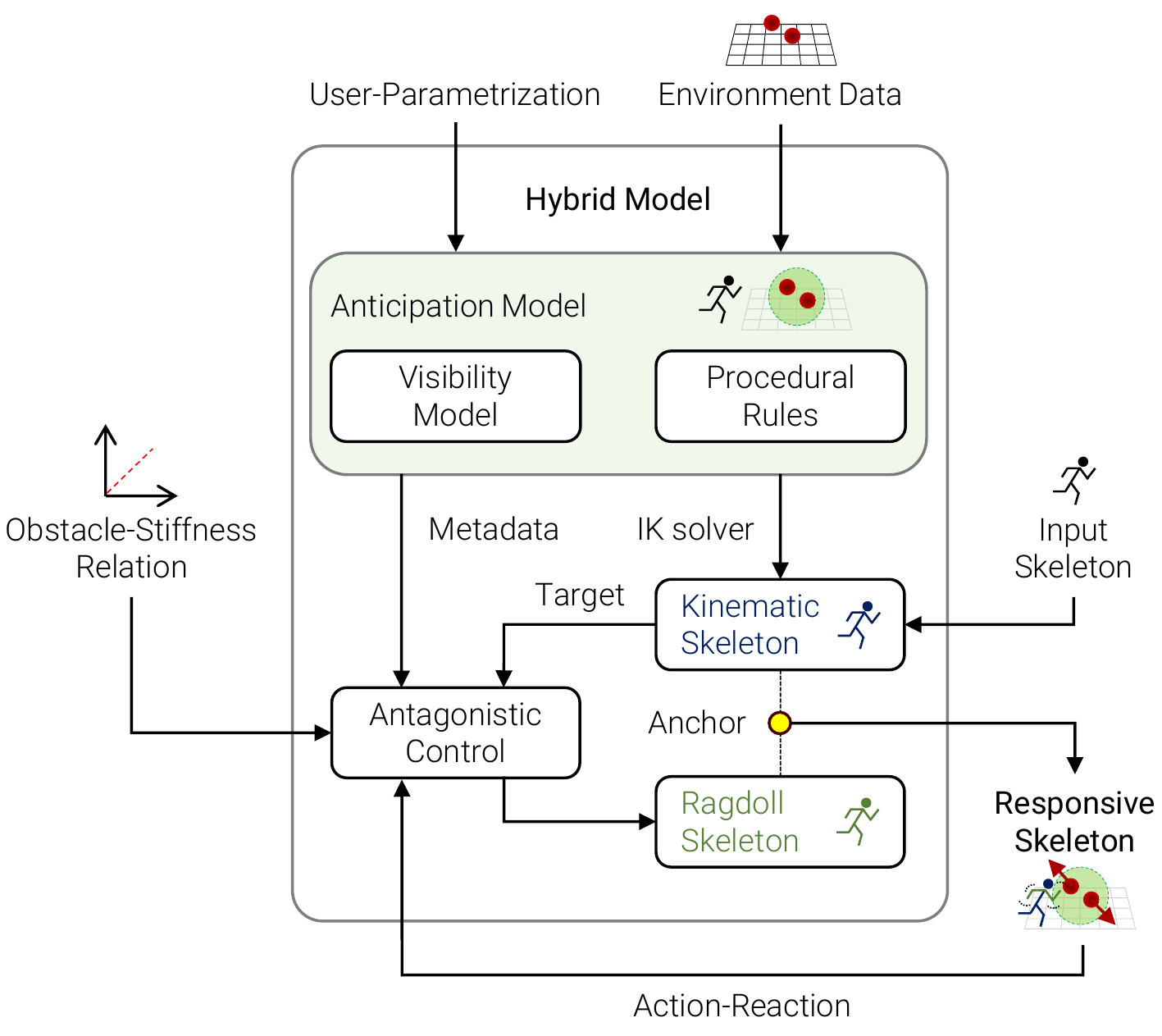}
  \caption{\label{fig:system-diagram} Overview of the animation pipeline for the hybrid model. 
  }
\end{figure}

Figure~\ref{fig:system-diagram} gives an overview of our animation pipeline. 
At each time $t$, we update the \textit{input skeleton} regardless of the environment. 
Then, we consider the local surroundings of the character, thanks to a lightweight visibility model.
Each obstacle type, 
its proximity, and its velocity are interpreted using a high-level rule-based system that provides 
kinematic objectives to the character's upper-body, such as pushing obstacles away with one or 
both hands, or protecting itself. These objectives are handled using an IK solver, and 
leading to the 
generation of an intermediate \emph{kinematic skeleton}. This skeleton can address high-level goals but does not integrate yet any dynamic response. To this end, we compute 
actuator torques
on 
a specific subset of dynamic limbs defined by an \textit{anchor}, using an antagonistic controller representation.
%
The latter allows the 
physical model to move
towards the time-varying kinematic skeleton, 
whatever our choice of stiffness (i.e., tension/relaxation) in the limbs.
This stiffness 
is adapted in real time, either from our anticipation model before 
establishing contact with an obstacle,
or from the current interaction forces
during contact. 
As a result, the target motion defined by the kinematic skeleton drives the physically-based limbs selected by the anchor system, leading to the final \emph{responsive skeleton} 
able to act on its dynamic surroundings, and to react to interactions in a plausible way.



\begin{figure}[htb]
    \centering
    \includegraphics[width=\linewidth]{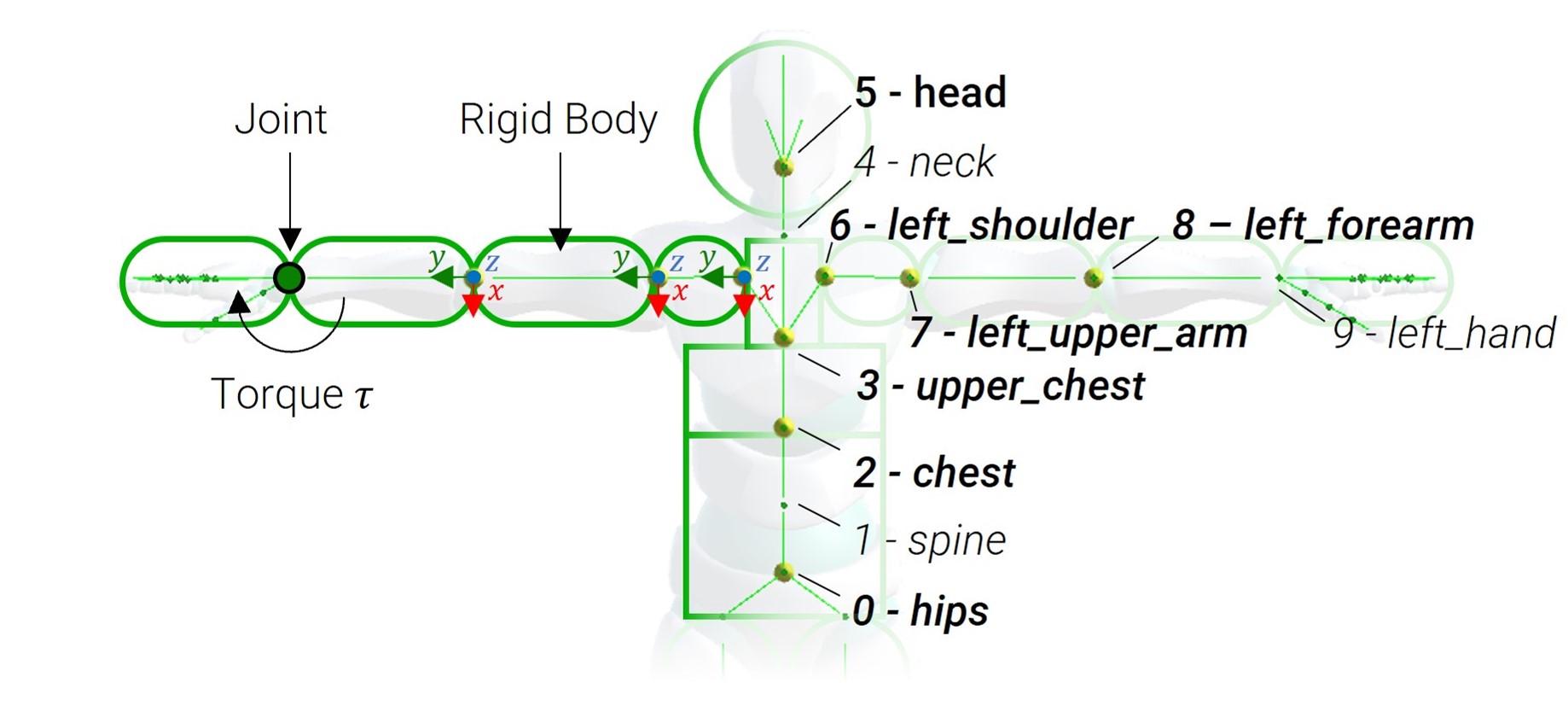}
      \caption{\label{fig:upper-body-rig} 
      Physical model of 
      a character. The anchor is chosen among the possible 
      positions \(a \in A \), shown as 
      spheres at the joints.
      The anchor $a$ partitions the skeleton hierarchy into a set of kinematic versus dynamic bones. 
      }
    \centering
\end{figure}

\begin{figure*}[htb]
  \centering
  \includegraphics[width=\linewidth]{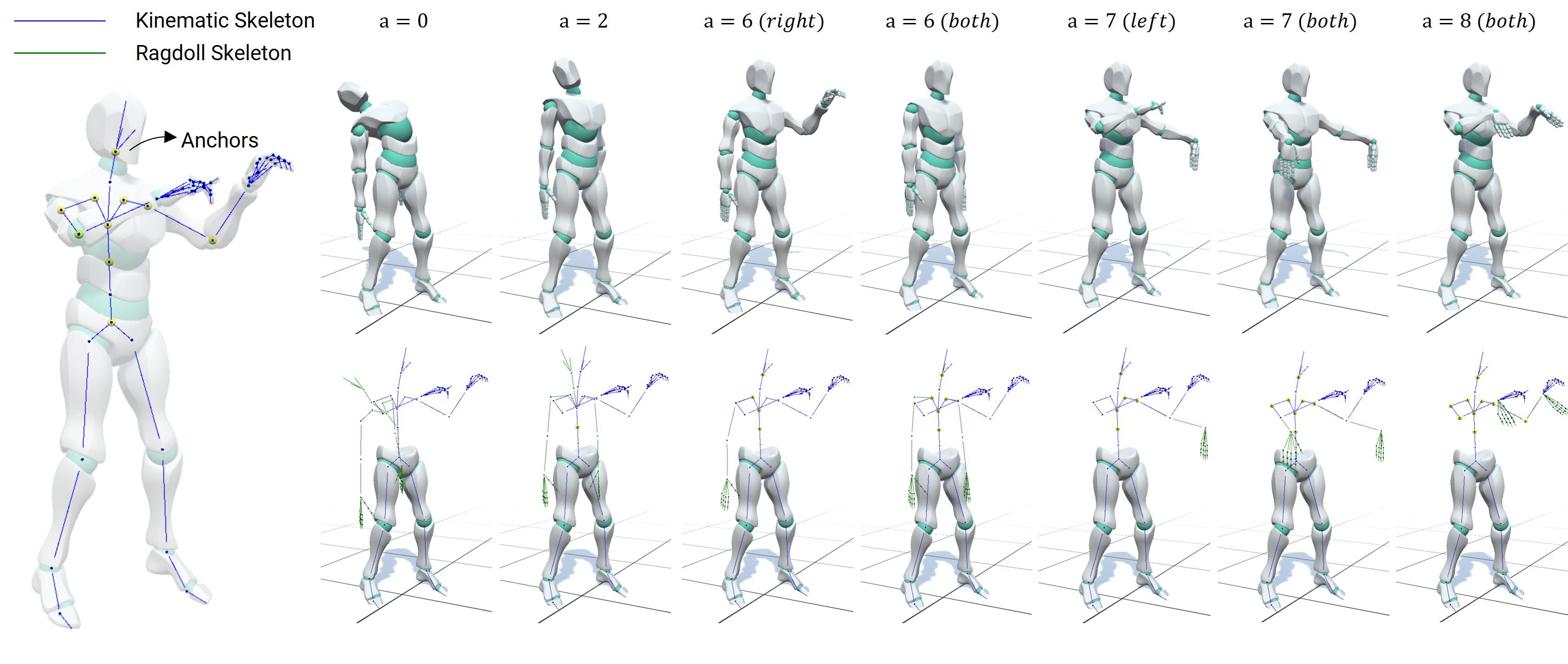}
  \caption{\label{fig:anchor-system-results} 
  Impact on the choice of an anchor on the responsive skeleton. Left: Input kinematic skeleton with yellow spheres representing the possible choices for an anchor.
  Right: responsive skeleton, where the weights of the limbs are only the external forces. The physical simulation takes control of the part of the character down the hierarchy, starting at the user-defined anchor point (where $a=0$ is the hips join, and $a=8$ is the wrist).
  The user may set an anchor on a single arm, or on both
  (shown for $a=6$ and $a=7$).}
\end{figure*}

We consider the following conventions to represent our skeletons (see Fig.~\ref{fig:upper-body-rig}): 
Each skeleton joint frame, located at the base of a bone, is encoded as 
a position~$p$ and an orientation~$q$ (unit quaternion), both w.r.t. its parent frame 
in the skeleton hierarchy. 
The local $Y$ axis is assumed to be aligned with the bone. In order to limit the physically-based simulation to a local subset of the skeleton, the final upper-body \textit{responsive skeleton} is 
partitioned into a set of kinematic parts and dynamic parts. Assuming that joint~$0$ corresponds to the root at the level of the hips, 
the partition between kinematic and dynamic bones is defined by 
the \emph{anchor}~$a\in A$, where $A$ is a set of admissible bones indicated in bold in Fig.~\ref{fig:upper-body-rig}. For a 
joint~$a$ and parent joint~$p(a)$, all ancestors 
$((p_0^{kin},q_0^{kin}),\cdots,(p_{p(a)}^{kin},q_{p(a)}^{kin}))$ are considered as kinematics-driven, while the descendants $((p_{a},q_{a}),\cdots,(p_{n},q_{n}))$ up to the end-effector~$n$ of the hierarchy are considered as dynamic 
ones with positions/orientations computed from the rigid-body simulator. 
Anchor $a$ plays the role of a local physical-root, and the choice of $a$ in the hierarchy depends on the nature of the current interaction.
Fig.~\ref{fig:anchor-system-results} 
shows the effect of choosing different nodes
for $a$ for a 
static character with no external, upper-body 
interactions, where only the weights of the limbs
are applied to the rigid-body simulation. Note that several anchors can be set at the same time, for instance for having both left and right arms be driven by dynamics, while the rest of the body remains fully kinematic.

%% file: Text/05-Antagonistic-Control.tex
\section{Extension of Antagonistic-based Control}
\label{antagonist_control}


\begin{figure*}[h!]
  \centering
  \includegraphics[width=0.9\linewidth]{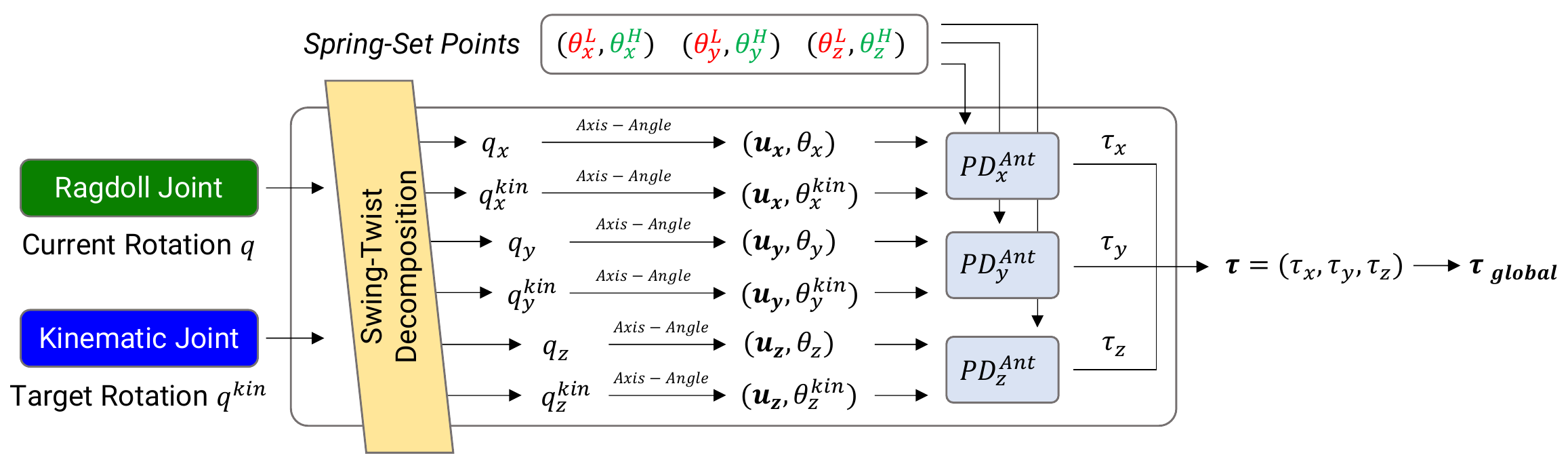}
  \caption{\label{fig:antagonistic-system} Antagonistic-based control for a 3-DOF joint. The system retrieves both, target and current orientation, and uses Swing-Twist Decomposition to get an orientation for each joint axis. Then, we convert the orientations to Axis-Angle representation and estimate the angular differences with respect to the user-defined spring-set points. Finally, these angular deviations are fed to the antagonistic controllers, which provide a torque that drives the physically-based model to the kinematic target orientation. 
  }
\end{figure*}

In this section, we first remind the general principle of antagonistic controllers~\cite{Neff2002} and their interest to control the dynamic part of the \emph{responsive skeleton}, before describing our formulation, designed for joints with two or three degrees of freedom.

\subsection{Antagonistic controllers principles}

Controllers for physically-based character models generally rely on Proportional-Derivative (PD) controllers, which convert the angular error in their proportional part to a spring-like force of prescribed stiffness.
However, on the one hand, setting a fixed value for the stiffness does not allow a skeleton to reach precisely a target orientation when 
external torques are applied, such as the effect of weight. On the other hand, changing the stiffness by increasing or decreasing it over time to reach an objective angle affects the style of the motion, by making it more or less tense or relaxed.
Derived from Feldman's theory on bio-mechanical motor control~\cite{Feldman1966}, the notion of antagonistic controller provides an elegant solution to stiffness control. First, it guarantees to reach 
the equilibrium at any arbitrary target orientation, specified within admissible bounds.
Second, it preserves the joint tension, and therefore the motion style, throughout the animation.
While antagonist controllers were already introduced in CG~\cite{Neff2002}, 
the method was developed for pre-computed target motions only, and the original approach suffered from gimbal lock issues when dealing with the elbow joints. 
We therefore propose a new formulation,
compatible with interactive motions where the target objective may change at run time, and expressed in local coordinates 
in order to avoid gimbal-lock issues.

Inspired from human anatomy, an antagonist controller models the action of a pair of antagonist muscles controlling the angle between two limbs at a given joint. 
The combined effect of two such muscles enables to reach any target angle with variable muscular stiffness.
Considering a single rotational degree of freedom and the relative angle \(\theta\) between the limbs 
with respect to the initial T-pose, 
the dynamics of the antagonist controller is parameterized by a pair of lower and upper stiffnesses
(or proportional gains)
\((k_L,k_H)\) associated to two angular soft limits
 \((\theta_L,\theta_H)\) 
at the joint,
and a derivative gain \(k_d\) such that
\begin{equation}
\tau + \tau_{ext}= k_{L}(\theta^{L} - \theta ) + k_{H}(\theta^{H} - \theta ) - k_{d} \dot{\theta} \;\; ,
\label{eq:antagonist}
\end{equation}
where \(\tau\) is the current actuator torque exerted by the controller and \(\tau_{ext}\) is the torque resulting from the external forces.

Let the motion of the articulation be guided by an objective target angle \(\theta^{kin}\), obtained from the kinematic skeleton.
When \(\theta\) reaches \(\theta^{kin}\), the 
antagonistic controller is at static equilibrium 
%
%
(\(\tau=0\)), leading to:
\begin{equation}
    \tau_{ext}^{kin} =  k_{L}(\theta^{L} - \theta^{kin} ) + k_{H}(\theta^{H} - \theta^{kin} ),
    \label{eq:equilibrium_controller}
\end{equation}
where \(\tau_{ext}^{kin}\)
is the total torque in this state.

\begin{figure}[htb]
  \centering
  \includegraphics[width=\linewidth]{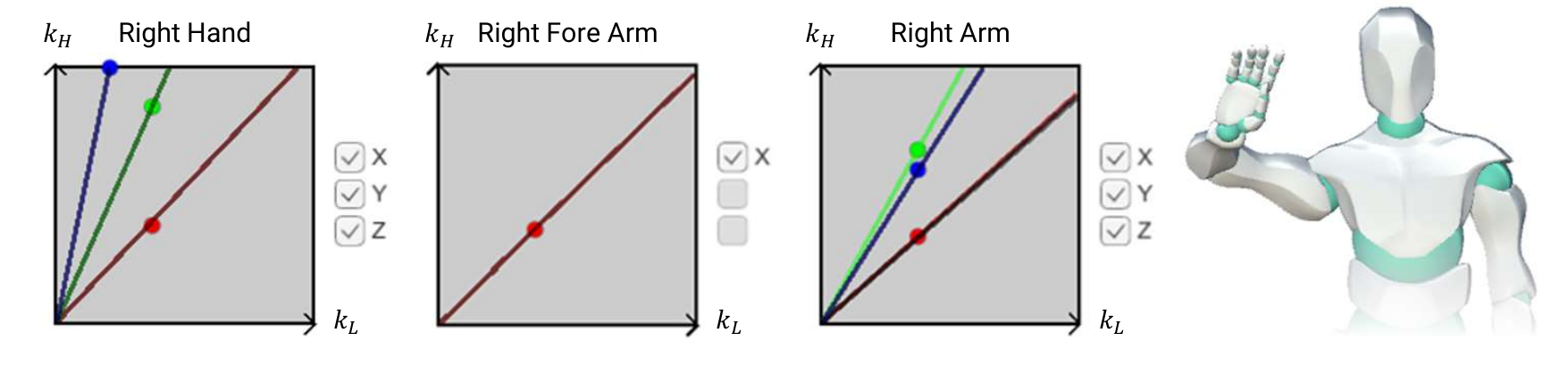}
  \caption{\label{fig:isolines} Isolines between antagonist stiffnesses, for each limb 
  of the responsive skeleton of the right arm.
  The slopes of the depicted line-segments are
  function of the  
 target angular pose.
   The isolines represent the degree of freedom of the stiffness to reach this specific pose. Navigating interactively along these lines allows an intuitive way to setup the stiffness values at the design stage of a game. At run time, our system further 
   computes automatically the corresponding coordinates along these lines based on a procedural rule taking into account the parameter of the obstacles (see Sec.~\ref{anticipation}).
  }
\end{figure}

This equilibrium constraint implicitly links the two stiffness values, and can be rewritten as:
\begin{equation}
    k_H = k_L \frac{\theta^L-\theta^{kin}}{\theta^{kin}-\theta^H}-\frac{\tau_{ext}^{kin}}{\theta^H-\theta^{kin}} .
    \label{eq:kh}
\end{equation}
This relation describes the so-called isoline of the controller, whose slope is independent from the external torques exerted on the joint. 
In addition to compensating for the respective error in the orientation, each pair of gains \( k_{L} \) and \( k_{H} \) lying on the isoline represents all the possible relaxation/tension configurations that the joint may use while reaching
the same equilibrium orientation. 
This simple linear relation thus provides us with an intuitive way of modifying 
the target orientation for the character while still being able to tune 
the desired amount of tension/relaxation in our pose (see Fig.~\ref{fig:isolines}).

\subsection{Adaptation to joints with multiple degree of freedom}

Antagonistic controller must be formulated in relation to each individual degree of freedom of a joint as it depends on extremal values \(\theta^L\),\(\theta^H\). A naive approach to extend this formulation to a joint with two of three degrees of freedom would be to decompose the joint orientation along Euler angle representation. However, decomposing a joint orientation expressed in a global frame system using Euler angles would lead to gimble lock artifacts during the animation. To avoid these artifacts, we propose to express this decomposition in a local frame  
where typical human-like gestures
will be free from Gimble-lock issues (see Fig.~\ref{fig:antagonistic-system}).

Let us consider the initial T-pose of the character, and call \(b^0\) the unit quaternion representing the orientation of a joint-frame expressed with relative coordinates to the parent joint.
We further attach in this relative coordinate system an orthogonal basis $(u_x,u_y,u_z)$ associated to the three degrees of freedom of this joint, and associate for each of them a low/high angular limit $(\theta^L_{x/y/z}, \theta^H_{x/y/z})$.
At run time, the articulated joint has an orientation given by the unit quaternion $b$ expressed in relative coordinates, and $q=b\, \overline{b^0}$ represents the rotation of by this joint in local coordinates, with $\overline{b^0}$ being the unit conjugate quaternion of $b^0$.
Similarly, we consider the target rotation $q^{kin}=b^{kin}\, \overline{b^0}$, $b^{kin}$ being the target orientation also expressed in the relative coordinates system of its parent.
Then, $q$ (resp. $q^{kin}$) is decomposed as $q=q_z\,q_y\,q_x$ (resp. $q^{kin}=q_z^{kin}\,q_y^{kin}\,q_x^{kin}$) using three consecutive Swing-Twist-Decomposition~\cite{Dobrowolski2015} along the axes $u_x$, $u_y$, and $u_z$. As such, \(q_x\) represents a rotation around the axis $u_x$, and similarly for the others.
Converting these decomposed quaternions into an axis-angle representation leads to the three angles \((\theta_x,\theta_y,\theta_z)\) (resp. \((\theta_x^{kin},\theta_y^{kin},\theta_z^{kin})\)) corresponding to the local rotation along each individual degree of freedom.
From these angles, the $(x,y,z)$ components of the torque \(\tau\) can be computed in this local frame using Eq.~(\ref{eq:antagonist}), and converted back to the global coordinate system before being used in the rigid-body simulator. Algorithm 1 summarizes this process.


\begin{algorithm}[t]
  \begin{algorithmic}[1]
    \Input{current $ q $, target $ q^{kin} $, spring-set points ($\theta^L$, $\theta^H$)}
    \Output{torque $ \tau $}
    \Function{ComputeTorque}{$ q $, $ q^{kin} $, $\theta^L$, $\theta^H$}
    \State $q_z\,q_y\,q_x$ $\gets$ STD$(q)$
    \State $q^{kin}_z\,q^{kin}_y\,q^{kin}_x$ $\gets$ STD$(q^{kin})$
    \ForEach {$u$ $\in$ DOF}
    \State $\theta_{u}$ $\gets$ $q_{u}$
    \State $\theta^{kin}_{u}$ $\gets$ $q^{kin}_{u}$
    \State Compute $\tau_{ext}^{eq}$ in equilibrium
    \State Initialize $k_{L}$
    \State $k_{H}$ $\gets$ $k_L \frac{\theta^L_{u}-\theta^{kin}_{u}}{\theta^{kin}_{u}-\theta^H_{u}}-\frac{\tau_{ext}^{eq}}{\theta^H_{u}-\theta^{kin}_{u}}$
    \State $\tau_{u}$ $\gets$ $k_{L}(\theta^{L}_{u} - \theta_{u} ) + k_{H}(\theta^{H}_{u} - \theta_{u} ) - k_{d} \dot{\theta_{u}}$
    \EndFor 
    \State \Return $ \tau $
    \EndFunction
  \end{algorithmic}
  \caption{Antagonistic Control}
  \label{antagonistic-algo}
\end{algorithm}

\begin{figure}[h!]
    \centering
    \includegraphics[width=0.9\linewidth]{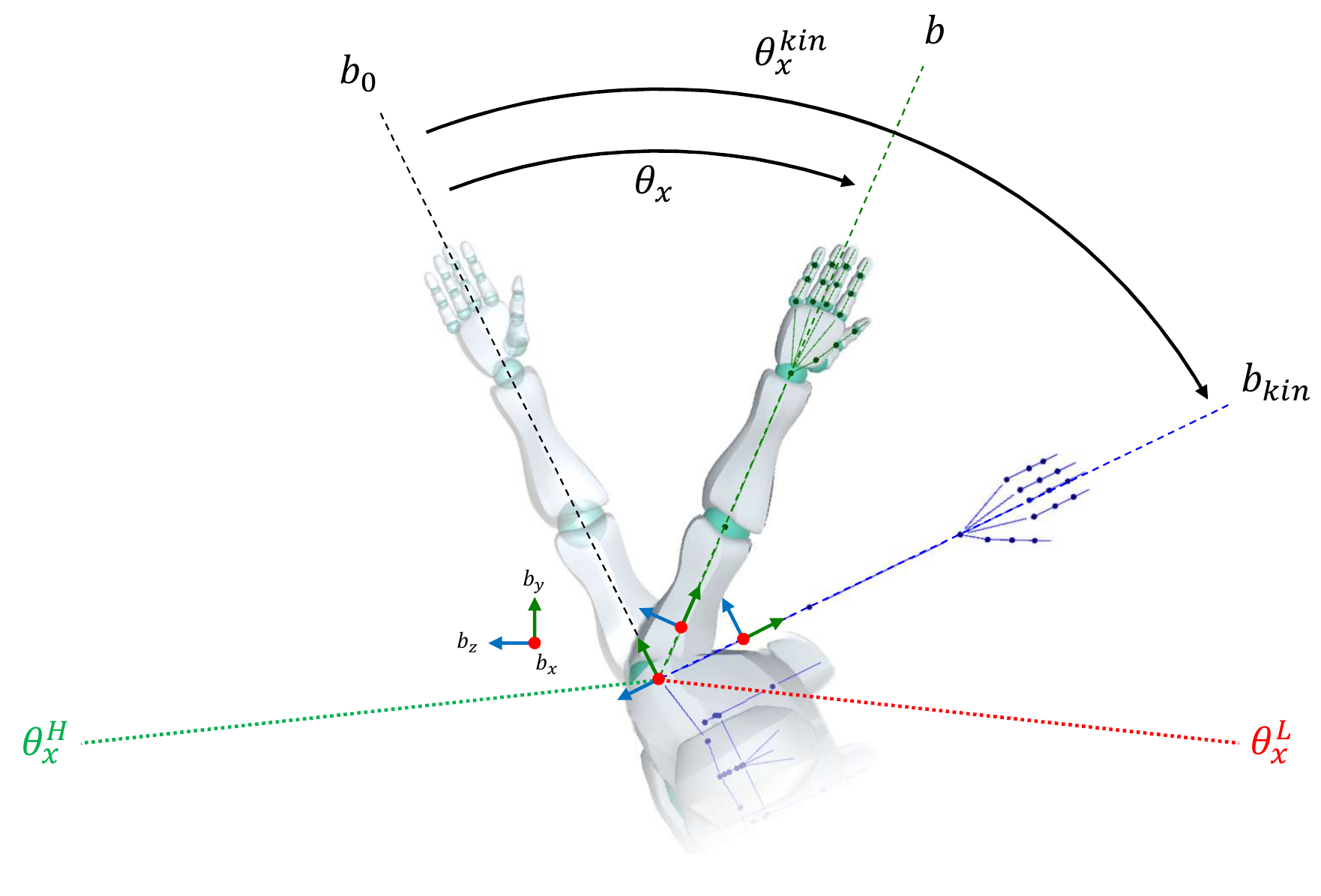}
      \caption{\label{fig:hand} 
      Left upper-arm configuration for the degree of freedom corresponding to the rotation around the local x-axis (vertical direction) at the shoulder. All angles are measured with respect to the initial T-pose orientation \( b_{0} \). \( \theta_{x} \) describes the current orientation, \( \theta_{x}^{kin} \) the target one, and \((\theta_x^H,\theta_x^L)\) are the high/low soft-limits of the antagonistic controller. 
      }
    \centering
\end{figure}

%% file: Text/06-Anticipation-System.tex
\section{Anticipation and Action on
Upcoming Obstacles}
\label{anticipation}



To free the user from manually tuning the responsive character every time it should interact with a new obstacle, we provide a high-level, yet customizable anticipation and action method, enabling the character to use its upper-body to make its way through complex dynamic environments. 
To realize this, an anticipation mechanism extracts metadata from the obstacles in the field of view, and uses it to edit the responsive skeleton model based on a set of 
procedural rules. 
These rules generate action gestures for the upper-body's kinematic skeleton so that the character protects itself and responds differently to obstacles of different nature.
They also automatically tune the antagonist stiffnesses of the ragdoll skeleton, so that obstacles can effectively be pushed out of the way.
In practice, this is achieved through one or several \textit{safety region(s)} around the character, which it will always try to keep free from obstacles; this is detailed next.

\subsection{Detecting Upcoming Obstacles}

\begin{figure}[htb]
    \centering
    \includegraphics[width=0.70\linewidth]{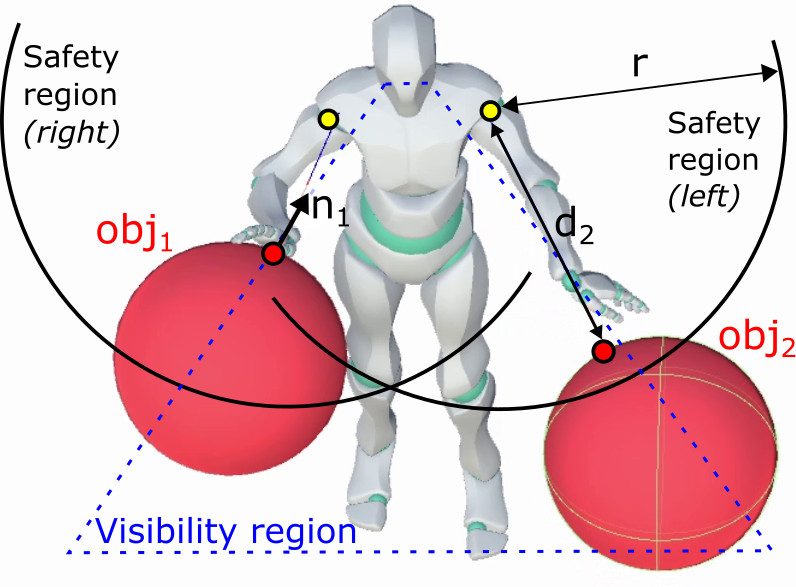}
      \caption{\label{fig:safety-region} 
      Visibility and safety Regions. The approximated frustum-like volume keeps track of the obstacles within the visual bounds of the character, while the safety regions are used to target the closest obstacles. 
      }
    \centering
\end{figure}

We use a frustum-like \emph{visibility region} starting from the character's head and propagating along the sight direction 
to make our character aware of his environment, and in particular of the obstacles to come.
We further define a set of \emph{safety regions} representing parts of the body that a character may try to protect with its arms while walking
in some dense, dynamic environment. We approximate these safety
regions as spheres around some body parts such as the left and right shoulders. Their radius $r$ can be defined from the character's size and arm length (note that using a slightly larger radius than the arm length is relevant, since it will enable the character to anticipate a collision by taking the right posture in advance, before eventually catching the object).
Our system considers that an obstacle must be stopped or pushed by the character's hand if it is both in the \emph{visibility region} and intersects with at least one of the \emph{safety regions }(see Fig.~\ref{fig:safety-region}).


\subsection{Anticipation behavior}

The character anticipates a possible future collision with the obstacles in actively pushing the one at the closest distance from the center of a safety region.
The choice of which arm to use to protect a particular part of the body is done according to which hand is at the closest distance to the obstacle, and the obstacle's mass: If the difference between the distances of the obstacle to each hand is less than a threshold, or if the estimated weight of the object is perceived as greater than a certain value, the character uses both hands to interact with it. 
Otherwise, when different obstacles appear at the left and right sides of the character, reactions from the two arms can be generated independently from each other, and may overlap in time.

The general idea of the anticipation gesture is to move the character's selected hand(s) to the closest point on the obstacle 
(or to the nearest accessible point to the obstacle, if the latter is not yet within reach),
while adapting the motion speed and tension to the expected velocity and mass of the latter. 
To this end, we provide a set of procedural motion rules aimed at generating somewhat natural behaviour, and which the user may customize to better reflect the specific personality of the character to be animated.
In addition, to model the possible uncertainty of the character's prediction, we allow metadata to return false or disturbed information about the expected mass and velocity of the upcoming obstacle. Considering an obstacle $obs_i$, $m_i$ and $v_i$ stand for the expected (possibly fake) mass and velocity, while $\hat{m_i}$, $\hat{v_i}$ are the actual ones.

Let us consider $p_0$, $q_0$ the original position and orientation of the arm at time $t_0$ when the obstacle became the targeted one. The closest point on the obstacle is defined by its position $p_c$ and associated normal $n_c$. We aim to orient the hand such that the palm becomes tangent to the obstacle at position $p_c$. We thus define $q_c$, the objective orientation, as the rotation transforming the direction \(b_x\) (orthogonal to the hand) into $n_c$, and the
$b_y$ direction (aligned with the hand) into $b_y \times n_c$ (see vectors conventions in Fig.~\ref{fig:hand}).
At a given time t, we consider the following kinematics objectives $p^{kin}$, $q^{kin}$ for the hand trajectory:
\begin{equation}
\begin{array}{l}
    p^{kin}(t) = p_0 + (p_c-p_0)\,\omega\left(\frac{t-t_0}{t_r}\right)  \\
    q^{kin}(t) = \textrm{SLERP}\left(q_0, q_c,\omega\left(\frac{t-t_0}{t_r}\right)\right)
    \label{eq:kinematics_objective_interpolation}
\end{array}
,
\end{equation}
where \(\omega\) is a smooth easing-function varying from 0 to 1, and $tr$ is the character's reaction time. 

We propose a simple 
way to set the character's reaction time, assuming that it only depends
on the expected relative velocity of the obstacle \(\|v_i\|\) with respect to the character's root's velocity:
\begin{equation}
    t_r = clamp\left(\alpha_{tr}\,\frac{d_i-r}{\|v_i\|},\, {t_r}_{min},\, {t_r}_{max}\right) \mbox{,}
    \label{eq:reactionTime}
\end{equation}
where \(d_i\) is the closest distance to the obstacle, $r$ is the radius of the safety region, \(\alpha_{t_r}\in[0,1]\) is a safety parameter ensuring that the hand should reach its final position slightly before the contact with the obstacle, and $({t_r}_{min}, {t_r}_{max})$ are user-defined bounds.

Note that the trajectory we just defined is, by construction, free of external obstacles, otherwise the hand would switch to another target position during the planned reaction time, to anticipate a more sudden collision.
In addition to these kinematics objectives, applied to either one or the two hands, 
the anticipation model also guides the stiffnesses $k_L$ and $k_H$ of the antagonist controller - which are linked by the linear law of Eq.~(\ref{eq:kh}). 
Indeed, we make the assumption that 
the character adapts his tension/relaxation behavior
based on the expected mass of the obstacle, leading to a relaxed motion for lightweight objects and more muscular tension when interacting with heavy ones. We model this behavior using the following linear relation between mass and stiffness:
\begin{equation}
    k_L=\mbox{clamp}\left(k_{L\,min} + (k_{L\,max}-k_{L\,min})\frac{m}{m_{max}},\, k_{L\,min},\, k_{L\,max}\right) \mbox{ ,}
\end{equation}
where $m_{max}$ is the extreme mass value that the character is expected to handle, and $k_{L\,min}$ and $k_{L\,max}$ are the limits for the lower gain $k_{L}$. The gain value $k_{H}$ is then computed accordingly using Eq.~(\ref{eq:kh}).

Note that the previously defined  bounds 
${t_r}_{min}, {t_r}_{max}, {k_L}_{min}, {k_L}_{max}$ are customizable.
The user can adapt them to a given character, but also, in all generality, set different bounds for different types of obstacles and safety regions. 
For instance, the character might always keep a slow motion and low stiffness when pushing away flexible and spiky vegetation such as brambles,
while being able to move faster and exert a higher tension toward other type of long-anticipated obstacles, such as a fence to be opened. Moreover, an obstacle moving towards the eyes of the character can thus be associated with a faster movement of anticipation than if it were moving towards his chest.

\subsection{Behavior during the contact phase}

\begin{figure}[htb]
    \centering
    \includegraphics[width=0.8\linewidth]{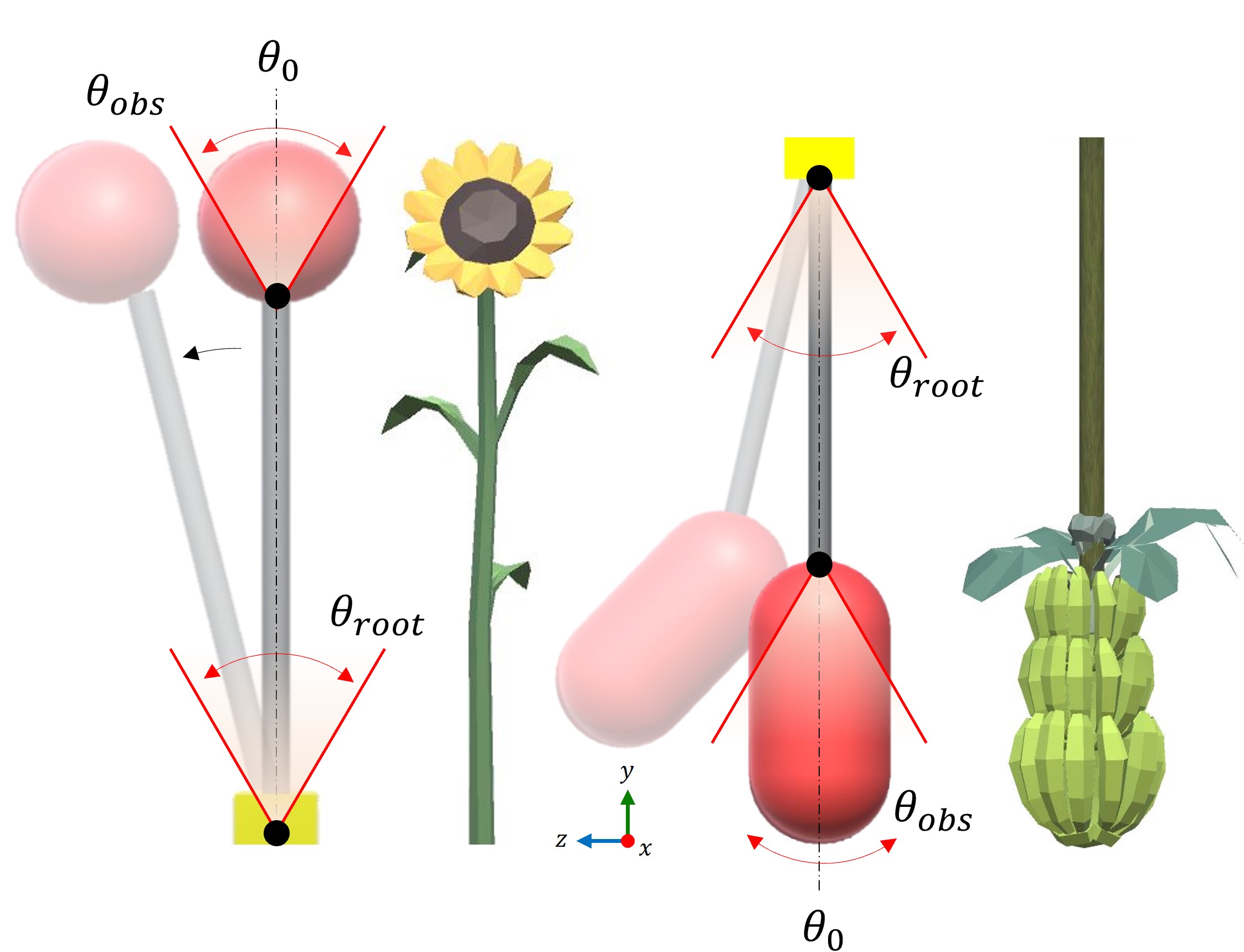}
      \caption{\label{fig:PD-assets} Obstacles such as flowers (left) or hanging fruits (right) are rigged and their background dynamics are defined by self-balancing poles that are kept upright using PD controllers. The red object (defined by the corolla, or by the fruits, respectively) 
      is used as a proxy for computing the interaction
      with the character.}
    \centering
\end{figure}

In our experiments, we consider the obstacles 
as dynamic articulated rigid bodies, visually represented as meshes deformed by skinning. The dynamic of these shapes is computed using a rigid body simulator, and the rest orientation of the articulated elements is handled using standard PD-controllers using a constant prescribed objective angle \(\theta_0\). 
Therefore, as long as the contact lasts between the hand and the obstacle, the opposite response forces generated by the collision are used to apply the corresponding external torques to the simulated character on one hand, and to the obstacle on the other hand (see Fig.~\ref{fig:PD-assets}).


\begin{figure}[htb]
    \centering
    \includegraphics[width=\linewidth]{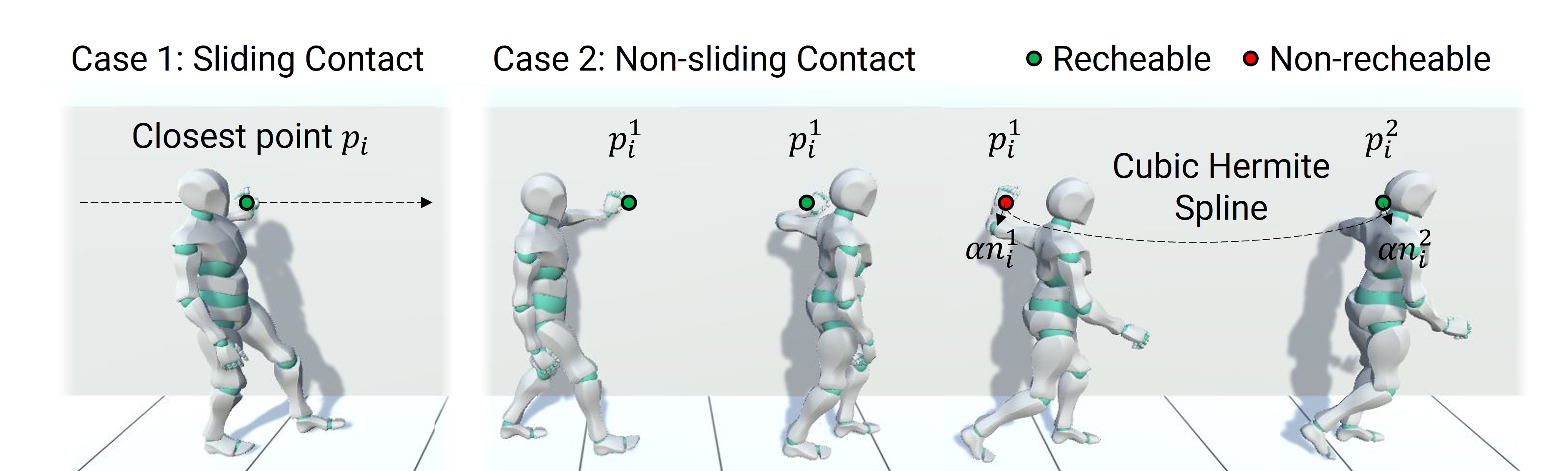}
      \caption{\label{fig:wall-scenes} 
      Depending on the user's choice, the character's interactions can be described as a continuous, sliding contact (left) or as a discretely updated contact position depending on the reach of the character's arms (right).}
    \centering
\end{figure}

In addition, our method can handle different kinematic-driven behavior for the character's arm, described as procedural rules. 
Our current implementation provides two specific scenarios:
The first one is when a character walks along a slippery obstacle along which the hand remains in contact but can slide.
To this end, the kinematic position of the hand is continuously adapted to target the updated closest point on the obstacle $p_i$ at the current frame, while remaining orthogonal to the normal at this position.
The second scenario holds in non-sliding contact cases (eg. contact with an uneven  wall), where the hand should remain at a fixed position \(p_i^1\) relative to the obstacle as long as the arm can reach this position. To do so, the original contact position is stored in the local reference frame of the obstacle, and used as the kinematic objective of the responsive skeleton, until the hand needs to be moved to a new position.
%
When this situation is detected, we trigger a temporary motion making the arm reaching a new updated closest point \(p_i^2\). 
During the transition period begin parameterized by the time $t_r$,
we consider a trajectory defined as a Cubic Hermite polynomial interpolating the two extreme positions $(p_i^1,p_i^2)$ with the two corresponding normals $(\alpha\, n_i^1, \alpha\, n_i^2)$, where \(\alpha>0\) controls how much the hand moves away from the obstacle.
During this transition, the orientation of the hand is interpolated using SLERP in the quaternion space. We illustrate these two scenarios in Fig.~\ref{fig:wall-scenes}.

Finally, when the targeted obstacle moves away of the safety region, the hand positions are interpolated, using a similar formulation than in Eq.~(\ref{eq:kinematics_objective_interpolation}), toward the next active obstacle in the priority list if it exists, or toward their current position in the 
kinematic skeleton otherwise.

\subsection{Failing at anticipating or handling interactions}

One of the key advantage of our method is that, as in real life, a character may miss-evaluate the nature of an obstacle or fail detecting it in time, and therefore not handle it properly. This makes the generated behaviour more lively. 
\begin{figure}[htb]
    \centering
    \includegraphics[width=0.9\linewidth]{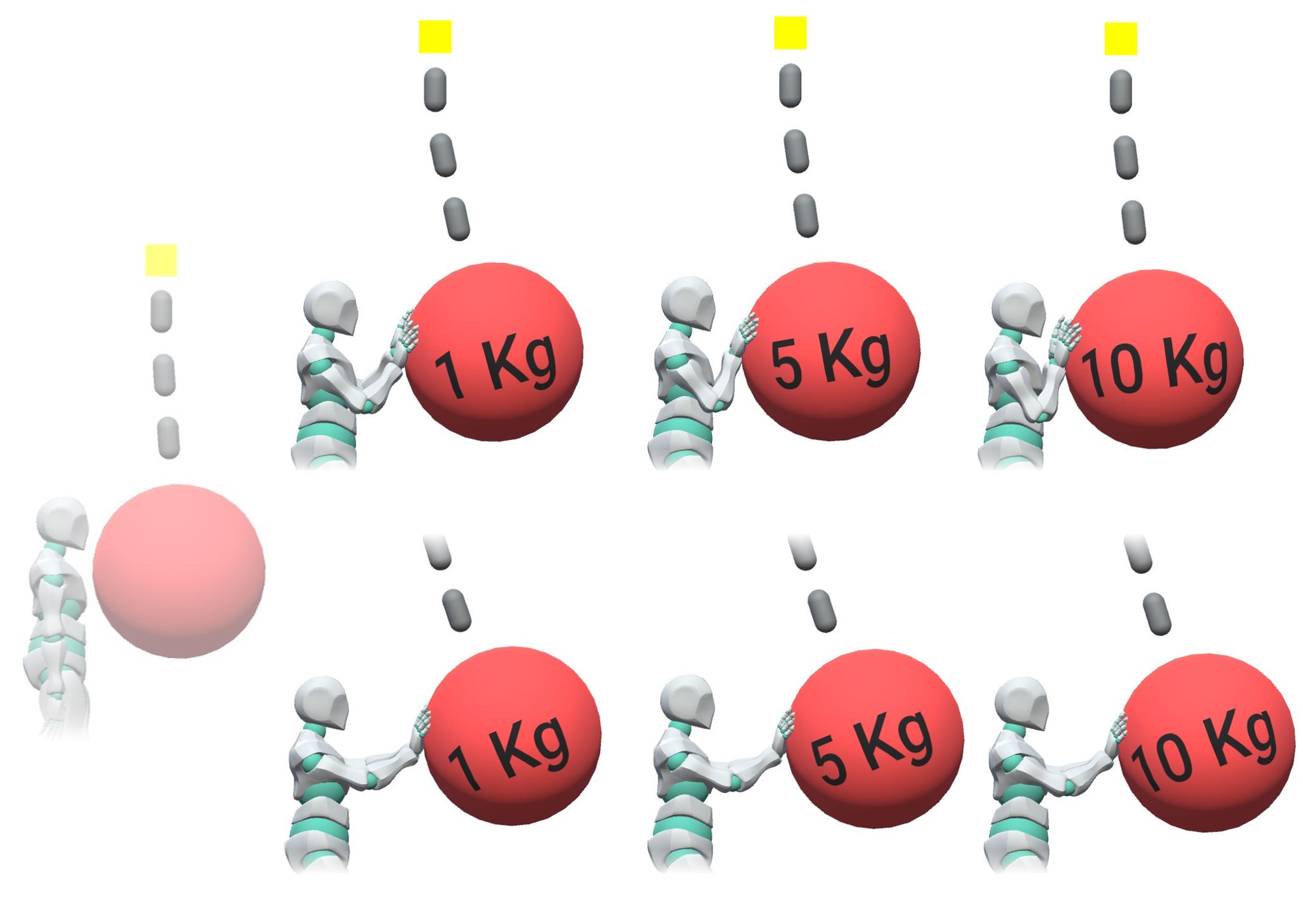}
      \caption{\label{fig:different-masses} 
      A character with constant tension in the arms (top) cannot maintain the prescribed posture when the mass of the red object increases, while an adaptation of tension (bottom) thanks to the antagonist stiffnesses in the controller enables to achieve it, within the limit of muscular strength.
      }
    \centering
\end{figure}

First, the stiffnesses of the antagonist controllers only takes into account the expected obstacle mass and velocity \((m_i,v_i)\),
which may be different from the real ones \((\hat{m_i}, \hat{v_i})\). As a result, a heavier or faster obstacle than expected leads to a controller experiencing large angular displacement in order to absorb the momentum of the obstacle as illustrated in Fig.~\ref{fig:different-masses}.  
Such miss-match will make the character look too \emph{relaxed}, and hardly able to avoid the collision. 

\begin{figure}[htb]
    \centering
    \includegraphics[width=\linewidth]{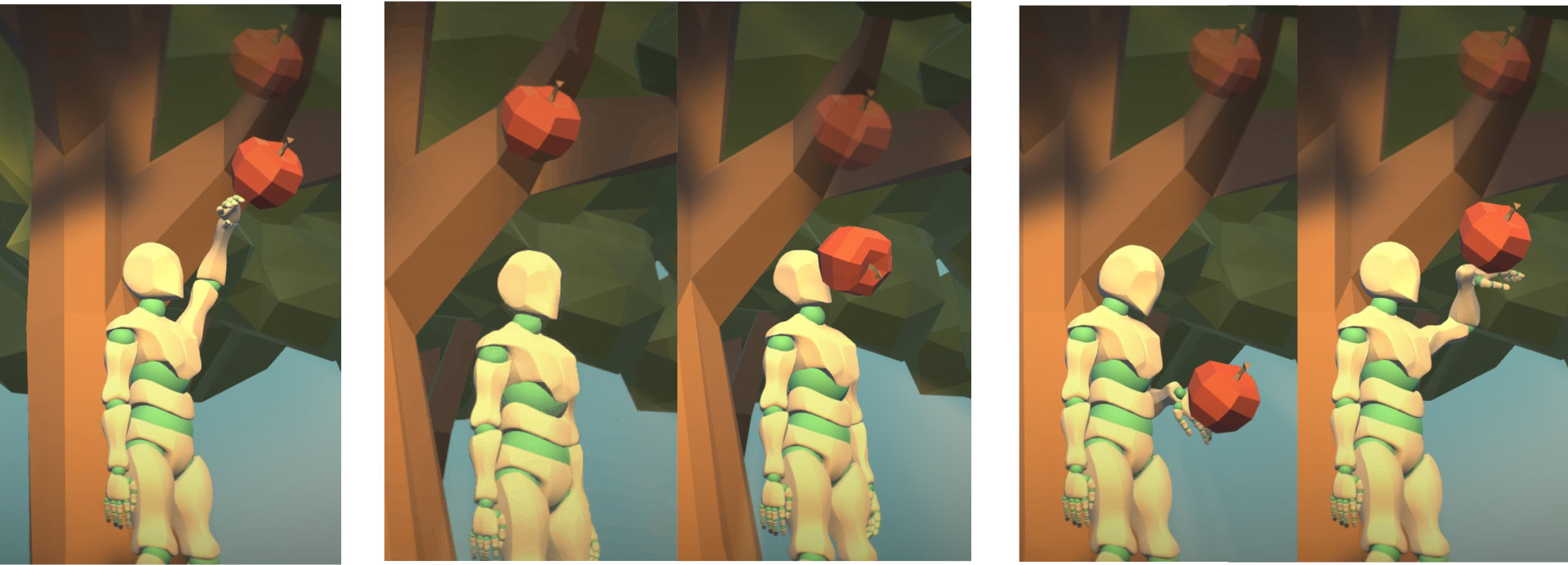}
      \caption{\label{fig:adapting-anchor} When the character 
      fails anticipating a collision,  
      our hybrid system system dynamically updates to generate an appropriate responds. Left: Expected correct handling of the falling apple. Middle: The apple fall outside of the visibility region leading to a collision with the head which is pushed back by the impact thanks to the dynamic update of the anchor point on the responsive skeleton. Then our antagonistic-based formulation helps it recover its initial position. Right: The apple falls too fast compared to the character's reaction time leading to non-optimal position to stop it. 
      }
    \centering
\end{figure}

When the expected mass is too low, reaction time may be too slow. Moreover, an obstacle coming from some unseen orientation may actually collide with the character without being detected. In this case, the ragdoll model is dynamically updated in order to model a reaction to an impact, while trying to restore the current position of the kinematic skeleton. Let us consider that the obstacle collides with a given limb. Then the closest possible anchor position on this limb or one of its parent is selected to be the root of a temporary antagonistic controller. We then adapt Eq.~(\ref{eq:antagonist}) in order to take into account the additional torque exerted by the obstacle, while the equilibrium condition from Eq.~(\ref{eq:equilibrium_controller}) is set with the angle at the impact time. As illustrated in Fig.~\ref{fig:adapting-anchor}-middle for a collision on the character's head, this approach allows to generate an adequate response to unexpected collisions as well.

%% file: Text/07-Results.tex
\section{Results and Discussion}

\begin{figure*}[h!]
    \centering
    \includegraphics[width=0.49\linewidth]{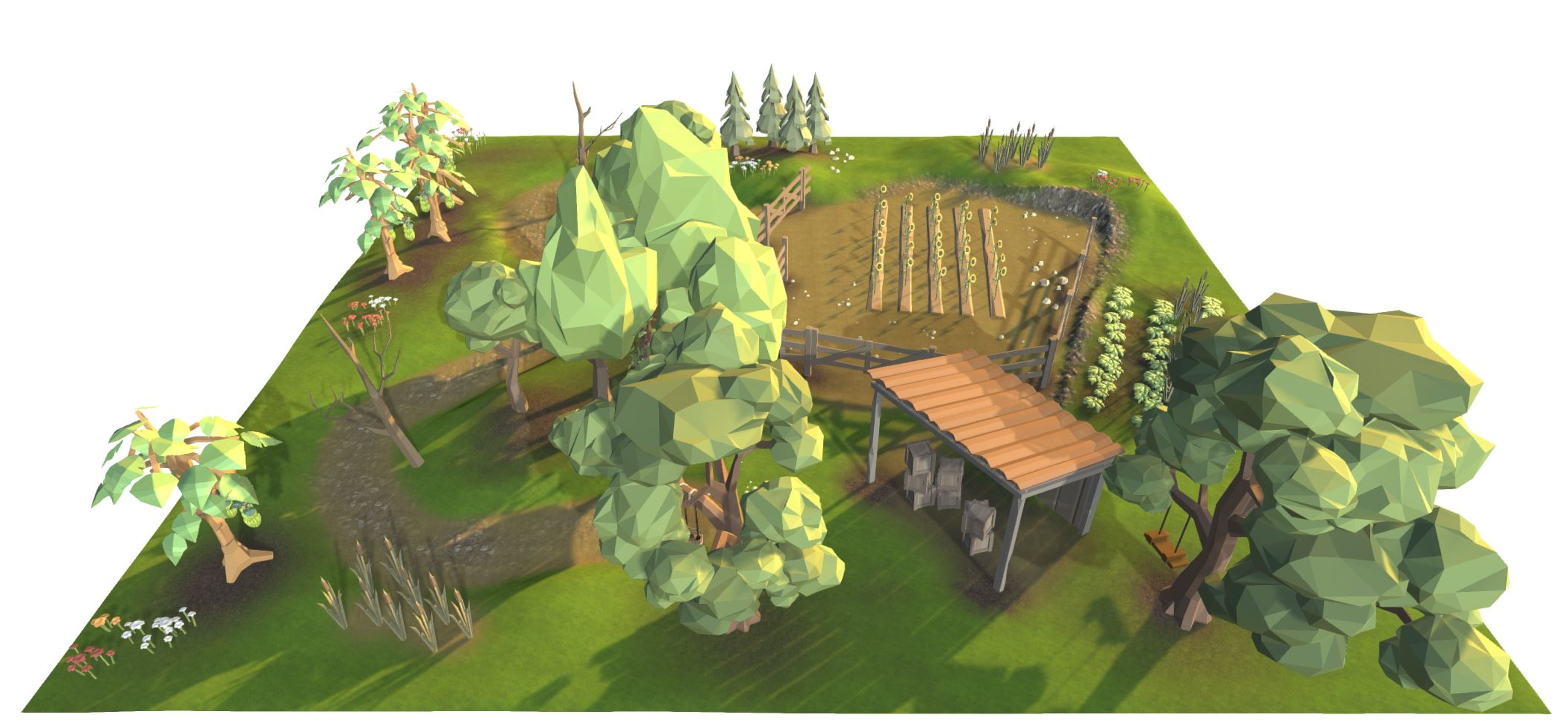}
    \includegraphics[width=0.5\linewidth]{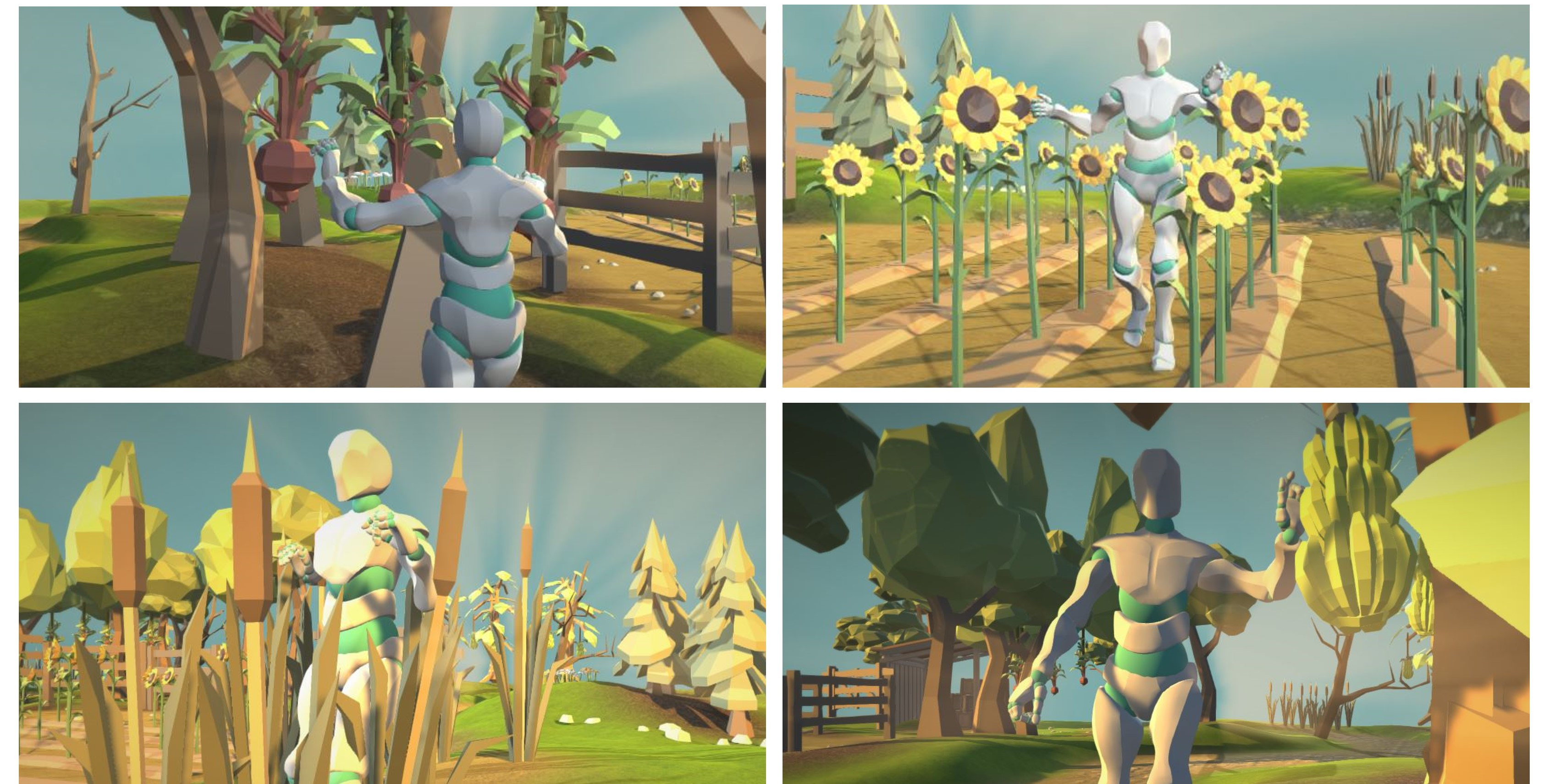}
      \caption{\label{fig:field-collage} Our test environment allows rich interactions with multiple dynamic elements, including flowers, flexible trees, a semi-rigid fence, and hanging objects. 
      }
    \centering
\end{figure*}

We implemented our method as an interactive prototype in the game engine \textsl{Unity 3D} where the character is interactively controlled using a game-pad or a keyboard. The entire method is coded in high-level C\# scripts,
and all the presented examples run in real-time on a standard laptop (\textit{Intel Core i7}, eight cores, running at 3.10~GHz). The main computational cost is spent on the rigid body simulation, while our additional anticipation model and procedurally-guided behavior do not bring any noticeable overhead. The default key-frame animation (with and without the IK) runs at 6 ms/frame (155 FPS). Activating the ragdoll simulation on the character with a single spherical obstacle adds an additional 1 ms/frame (140 FPS). Our most complex scene (Fig.~\ref{fig:field-collage}) includes 70 simulated plant assets and requires 25ms/frame (40 FPS). Note that these measures could be optimized in skipping the simulation of the assets that are not interacted with.

In our experiments, the character is placed in an environment consisting on various dynamic elements, such as different vegetation types or hanging objects. 
The input kinematic skeleton is animated using a state-machine controller with keyframe animation of a walking gait.
%
Antagonistic-based controllers are set at each degree of freedom of the shoulders, along with an initial stiffness
\( k_{L} \) and angular limits
\( \theta^{L} \) and \( \theta^{H} \). 
%
The character's upper-body is protected using two spherical safety regions of radius \( r = 0.5 \ meter \), located at the center of each shoulder joint. 
%
We use a default reaction time \( t_{r} = 0.5 \ s \) with minimum and maximum bounds of \( {t_r}_{min} = 0.5 \ s \) and \( {t_r}_{max} = 2 \ s \). 
The maximum mass that the character is expected to handle is by default \( m_{max} = 15 \ kg \). 
%
See Appendix~\ref{appendix} for additional information on the parameters used in our examples. 

The animated results, described next, are provided in the companion video.

\subsection{Interacting with Different Stiffnesses and Reaction Times}

Fig.~\ref{fig:teaser} illustrates a general situation where the character interacts with different types of outdoor elements. The diverse nature of the elements in terms of sizes or weights, are associated to different, but coherent, actions and reactions of the character.

A constant controller stiffness
would lead to a fixed amount of tension during this journey.
As such, the character would never adapt its muscular strength 
to the obstacle's mass, 
leading to increased bending of the arms and difficulty to push heavier obstacles away, as shown in Fig.~\ref{fig:different-masses}. 
Thanks to our local adaptation to the anticipated weight, from Eq.~(\ref{eq:kh}), the character is able to dynamically adapt to the current obstacle with a similar posture, in the limits of its muscular capacity.

\begin{figure}[htb]
    \centering
    \includegraphics[width=\linewidth]{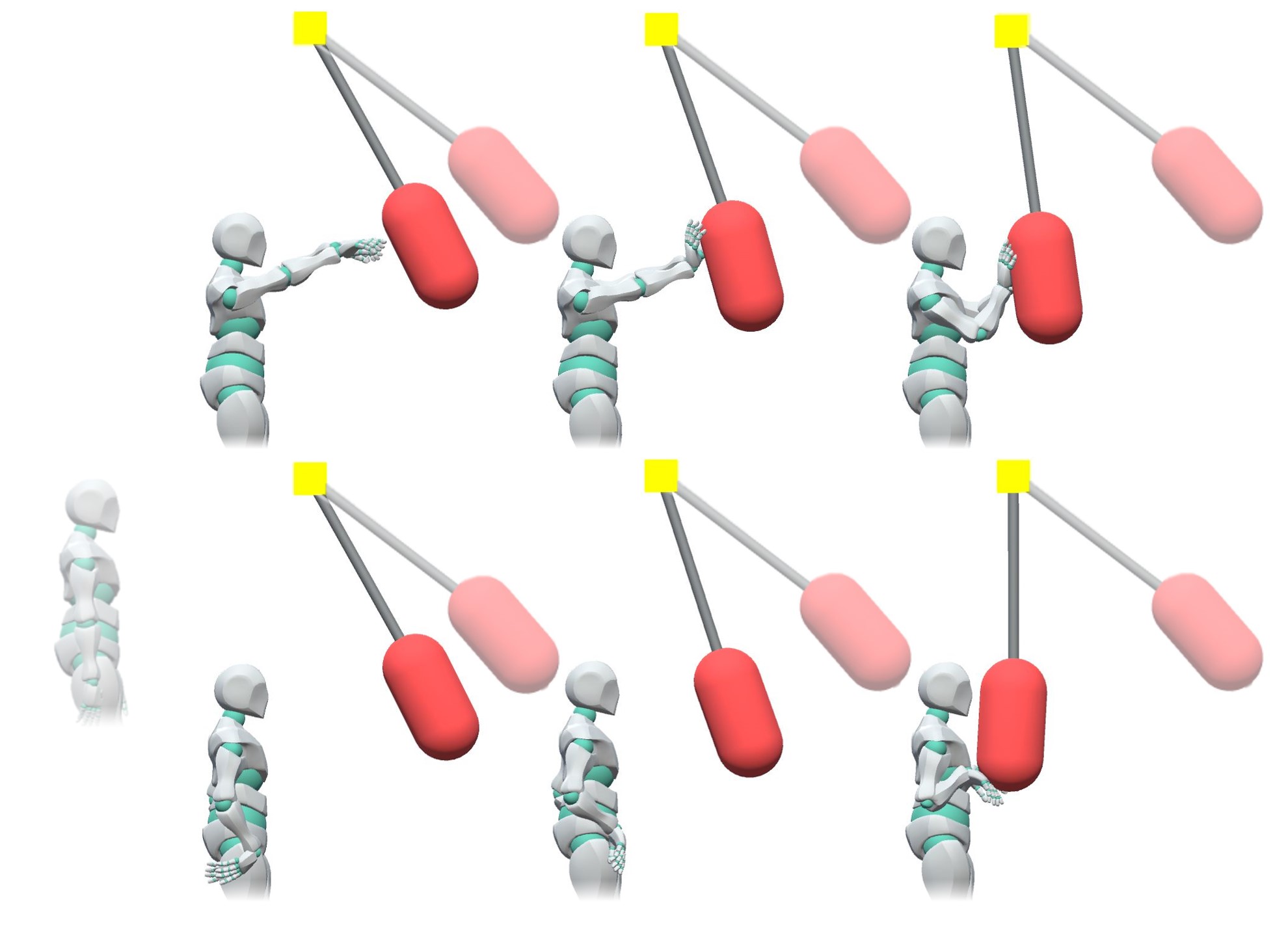}
    \caption{\label{fig:different-times} Short reaction time (top) vs longer (bottom), with detection time and all other parameters unchanged.
    Note that by using the longer reaction time, the character did not have time to properly place his hand before impact.
    }
    \centering
\end{figure}

The reaction time \( t_{r} \) may affect the success of an interaction in certain cases (see Fig.~\ref{fig:adapting-anchor}-right and Fig.~\ref{fig:different-times}). If the character reacts too slowly when an obstacle is heading its way, its arms will not take an optimal position to stop the obstacle. Reducing reaction time 
increases the speed at which the reactive skeleton assumes the 
correct protective posture, making it easier to manage the interaction.

\subsection{Anticipating Natural Surroundings}

Table~\ref{tab:summary-factors} summarises both internal factors and external metadata that can be retrieved by the anticipation system, along with their impact on the behaviour of the interaction. To better show these effects and experiment with interactions in dynamic environments, we built a 3D scene comprising multiple assets, as illustrated in Fig.~\ref{fig:field-collage}. Appendix~\ref{appendix} provides the set of parameters used for our 3D models.

\begin{figure*}[htb]
    \centering
    \includegraphics[width=0.65\linewidth]{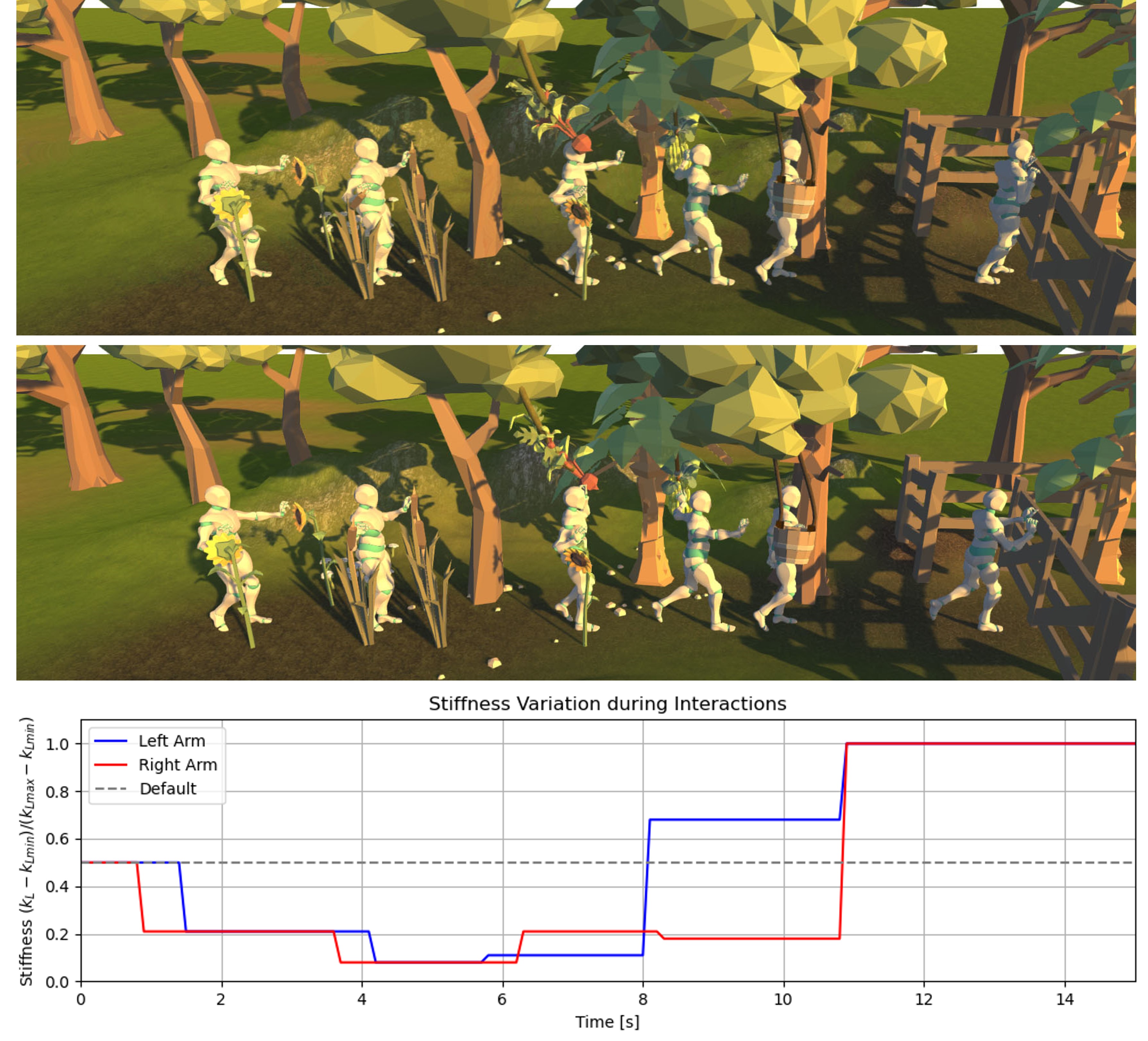}
      \caption{\label{fig:adapting-stiffness} 
      While light vegetation like tall grass can be easily pushed away, heavier items like the wooden gate may require more muscle stiffness to be pushed back. When the character does not adapt its tension/relaxation to obstacles (top), it results in over-stiff movements for light plants, inaccurate contacts, and over-relaxed motions when facing heavier obstacles. With our method (middle), the character automatically adapts the stiffness of each arm while making its way through. Bottom: Depiction of the stiffness variation along the obstacles.}
    \centering
\end{figure*}

We tested different types of behaviors dependent on metadata in this environment, using our high-level procedural rules for cases such as quick anticipation movements, or long-term behavioral responses. 
%
The character adapts its muscular rigidity based on the expected mass retrieved from the active obstacles entering the safety region. A comparison between character interactions with, versus without, such adaptation is shown in Fig.~\ref{fig:adapting-stiffness}, as well as in the companion video.
Moreover, since the velocity of an incoming object 
is used to adapt reaction time \( t_{r} \) (see Eq.~(\ref{eq:reactionTime})), the character is able to automatically generate quick anticipation movements to protect itself. 

\begin{table}[htb]
\centering
\begin{tabular}{l l} 
    \toprule
    \textit{\textbf{Internal Factor}} & \textit{\textbf{Description}} \\
    \midrule
    Safety Region Radius \( r \) & Reach of the anticipation \\
    Reaction Time \( t_{r} \) & Rapidity of the interaction \\ 
    \makecell[l]{Reaction Time Bounds \\ \( [{t_r}_{min}, {t_r}_{max}] \)} & Reaction capacity (faster/slower) \\
    Stiffness Gain \( k_{L} \) & Muscle rigidity \\
    \makecell[l]{Stiffness Gain Bounds \\ \( [k_{L\,min}, k_{L\,max}] \)} & Stiffness capacity (stronger/weaker) \\
    \toprule
    \textit{\textbf{External Factor}} & \textit{\textbf{Impact of Anticipation}} \\
    \midrule
    Expected Mass \( m \)  & Increase/Decrease Stiffness \( k_{L} \) \\
    Expected Velocity \( v \) & Increase/Decrease Reaction Time \( t_{r} \) \\
    \bottomrule
\end{tabular}
\caption{\label{tab:summary-factors}Internal parameters and external factors that influence character behavior. Note that both, expected the mass and velocity returned by the metadata, might differ from real values. The character will then mis-adapt its behaviour, according to its current perception of the object.}
\end{table}

\subsection{Discussion and Limitations}

The previous results illustrated the benefits of our method in various scenarios, through real-time interactions with a variety of obstacles. While they demonstrate the benefits of automatic tension/relaxation adaptation, thanks to our new formulation of antagonist control, the current implementation is limited to the case of characters pushing obstacles away from their safety regions. 
To improve realism, more complex behaviors could be integrated, such as refining our anticipation process with better predicted hazards, modeling protective or avoidance gestures for obstacles with large momentum, and dynamically adapting the character's response when the true mass of an obstacle is perceived, during interaction.
We believe that these would be quite easy extensions of our method, thanks to our flexible procedural approached, guided by the observed environment. 

As our prototype was developed as a generic proof of concept, some of the parameters and procedural rules should be refined to improve the plausibility of motions and contact modeling. For instance, 
the character's wrist sometimes has too great a twist angle during the anticipation phase or rotates too slowly in relation to the arm. 
This could be addressed via dedicated IK constraints for the wrist. Additionally, some small gaps can also be seen between the hands and the mesh used to represent the vegetation. The use of finer collision proxy representation would avoid such visible artifacts.
Also, the velocity (magnitude and direction) of the obstacles could be considered in the anticipation behavior to deal with multiple obstacles more plausibly, as our current system only handles the closest one.

The main limitation of our approach is obviously the independence between the upper-body motion and the orientation of the character, as well as the locomotion gait of the lower-body. Indeed, in real-life, the motion of our legs is linked to our actions. For instance, we stop moving when expecting an unavoidable hit, or we bend hips and knees in order to push heavy obstacles. Such coupling is not orthogonal to our approach, but would require - on top of additional rules - 
a more global simulation method including a locomotion controller.

Lastly, while enabling precise authoring of the character's behavior thanks to procedural rules is usually desired in video-game development, the number of cases to explicitly define can be a burden. In addition, some of the parameter choices may result in limited plausibility, and the rules themselves may have a limited validity range. For instance, defining the controller's stiffness as a linear function of the obstacle mass may not be a valid approximation anymore for a wide range of masses, or for objects moving at high velocities. 
This choice may be a limitation in some cases compared to more complex, learning-based approaches.
Using such methods could allow to tune tension/relaxation not only in anticipation before a collision, but also continually during the contact phase, resulting in more convincing interactions.








%% file: Text/09-Conclusion.tex
\section{Conclusion and Future work}

In this work, we have proposed a method to generate real-time upper body movements from simple keyframe locomotion inputs, such that characters make their way through dynamic environments. Our hybrid model for character animation combines kinematics, IK constraints, and lightweight physics to produce a responsive skeleton able to react to any kind of external obstacle. A local anchor system identifies the limbs that need to be dynamically simulated during the interaction, and leveraging antagonistic controllers to generate actuator torques that follow a reference animation, while controlling the level of tension/relaxation independently. A flexible anticipation mechanism allows the user to combine both, information from the surroundings and changes in the character's stiffness and reaction time, enabling high-level authoring in the way the character handles the interactions. 
Overall, we believe our reactive character model provides a practical and flexible framework well suited to video game pipelines where precise control of behaviors and real-time computation are essential.

A promising future direction would be to explore the coupling of our approach with some non-supervised learning method, in order to improve the fine-grain plausibility of interactions while maintaining the same level of control. Procedural creation could then be used to define the coarse behavior, while reinforcement learning would help guide and enrich the local details of the animation, thanks to the fine-tuning of the antagonistic control for each joint, set to minimize the muscular effort undergone by the character.

%% file: Text/99-Acknowledgments.tex
\section{Acknowledgments}
This work has received funding from the European Union’s Horizon 2020 research and innovation programme under the Marie Skłodowska-Curie grant agreement No 860768 (CLIPE project). We also thank Prof. Pierre Poulin (Université de Montréal) for his help and comments on the manuscript.

%% file: Text/10-Appendix.tex
\section{Appendix}
\label{appendix}

\subsection{Character and Environment Description}


Table~\ref{tab:rigid-bodies} gives the parameters used in our experiments for each rigid-body in the responsive skeleton.

\begin{table}[htb]
\begin{tabularx}{\linewidth}{c*{8}{c}} 
    \toprule
    \textit{\textbf{Rigid-body}} & \textit{\textbf{Mass (kg)}} & \multicolumn{3}{c}{\textit{\textbf{Center of Mass (x,y,z) (m)}}} \\
    \midrule
    Hips\&Spine &   10.56   &    0.00         &   0.00    &   0.00       \\
    Chest       &   25.2    &    0.00         &   0.22    &   -0.03      \\
    Upper Chest &   10      &    0.00         &   0.35    &   -0.04       \\
    Neck\&Head  &   4.8     &    0.00         &   0.60    &   0.00       \\
    Shoulder    &   1       &   $\pm$ 0.06    &   0.44    &  -0.04     \\
    Upper Arm   &   2.95    &   $\pm$ 0.19    &   0.42    &  -0.03      \\
    Forearm     &   1.59    &   $\pm$ 0.23    &   0.15    &  -0.02     \\
    Hand        &   0.5     &   $\pm$ 0.26    &   -0.13   &   -0.02     \\
    \bottomrule
\end{tabularx}
\caption{\label{tab:rigid-bodies}Rigid-body settings for the upper-body model.}
\end{table}


Most of the 3D obstacles in our environments (see Fig.~\ref{fig:PD-assets}) are rigged and defined using PD controllers with different rest-positions. Therefore, both fences and hanging plants follow similar principles, each set to a different dynamics. The main parameter values are given in Table~\ref{tab:natural-assets}.

\begin{table}[htb]
\centering
\begin{tabular}{c c c c} 
    \toprule
    \textit{\textbf{Asset}} & \textit{\textbf{Mass (kg)}} & \textit{\textbf{$k_p$}} & \textit{\textbf{$k_d$}} \\
    \midrule
    Sunflower (Large)    &   1.25   &   20      &    10     \\
    Sunflower (Small)    &   1      &   20      &    10     \\
    Bush                 &   0.2    &   200     &    10     \\
    Banana Tree          &   10     &   50      &    1      \\
    Tree Branch          &   3      &   500     &    15     \\
    Fence                &   10     &   100     &    10     \\
    Fence w/ Door        &   10     &   100     &    20     \\
    Hanging Bucket       &   1      &   -       &    -      \\
    Swing                &   2      &   -       &    -      \\
    Hanging Fruit        &   0.5    &   -       &    -      \\
    \bottomrule
\end{tabular}
\caption{\label{tab:natural-assets}Obstacles used in our natural environment.}
\end{table}